# Fractional Separation of Polymers in Nanochannels: Combined Influence of Wettability and Structure

Sree Hari P D[1], Chirodeep Bakli[1], Suman Chakraborty[1]

[1]Department of Mechanical Engineering, Indian Institute of Technology Kharagpur, Kharagpur 721302, INDIA.

Correspondence to: Suman Chakraborty (E-mail: suman@mech.iitkgp.ernet.in)

**ABSTRACT**

Trapping macromolecules in nanopits finds multifarious applications in polymer separation, filtering biomolecules etc. However, tuning the locomotion of polymers in channels of nanoscopic dimensions is greatly restricted by the comparative advective and diffusive components of velocities. Using the polymer affinity toward the solvent and the wall, and the polymer structure, a mechanism is proposed to induce selective trapping of polymers. Similar to fractional distillation of hydrocarbons based on molecular weight, a technique of fractional segregation, depending on the channel wettability of polymeric chains at different depths in a pit that is located perpendicular to the flow is suggested. Depending on the properties of the polymeric chains and the surface chemistry, the segregation of the polymer at a particular level in the pit can be predicted. This behaviour stems from the difference in polymer structure leading to a competition between wettability based trapping and entropic trapping. The results of this study suggest a novel way of separating biopolymers based on their structure without relying on the channel geometry.

**INTRODUCTION**

Detection and sequencing of bio-polymers like DNA[1,2], proteins and other organic molecules and a plethora of industrial processes have become closely associated with nanoscale technology.[3–9] The strong interfacial forces over the volumetric forces, over these scales, help to achieve increased sensitivity in innumerable pharmaceutical and bio-medical applications.[10–13] Interestingly, polymers can have extremely disparate length scales depending on the degree of polymerization and hence widely varying time scales. Since, it is difficult to manoeuvre the classical flow properties at these scales, it is advantageous to manipulate size-selectivity depending on length scales and flow rates. Confinement with complex geometries can bring in free energy gradients in the system and consequently macromolecules like DNA will undergo entropic trapping.[14–16] The phenomena of entropic trapping and recoiling, as confirmed from experiments as well as simulations, [17,18] aid to understand and design models to separate and sort complex molecules like DNA.[14,15,19–21] Moreover, polymers in solvated state like water, are expected to give rise to non-intuitive behaviour.[22–27] Irrespective of the copious amount of literature existing on the detached problems: charge-wettability interaction in molecular fluids and polymer dynamics at nanoscales, few attempts had been made to utilize the channel wettability, charge interactions and polymer structure in confinement toward filtration and segregation of polymers. Such mapping of selective transfer of polymers can enhance the understanding of migration of reactive species and signals in several natural processes within cellular environment.

The present study differs from the preceding studies of polymer filtration in two major aspects. Firstly, the dimensions of the channel and the pit are of the same order and polymeric chain dimensions are much smaller. This excludes the contribution to polymer trapping from chain relaxation. In other words, the filtration process would be independent of the channel geometry and hence, precise nanofabrication techniques would not be required. Secondly, we include three modes of interaction: the polymer-solvent, solvent-wall and wall-polymer, as opposed to considering only a solvated polymer. This consideration not only makes the simulation system more physically realistic, but also enables us to use the properties of the polymer themselves to alter its flexibility and trap it at a particular level of the pit.

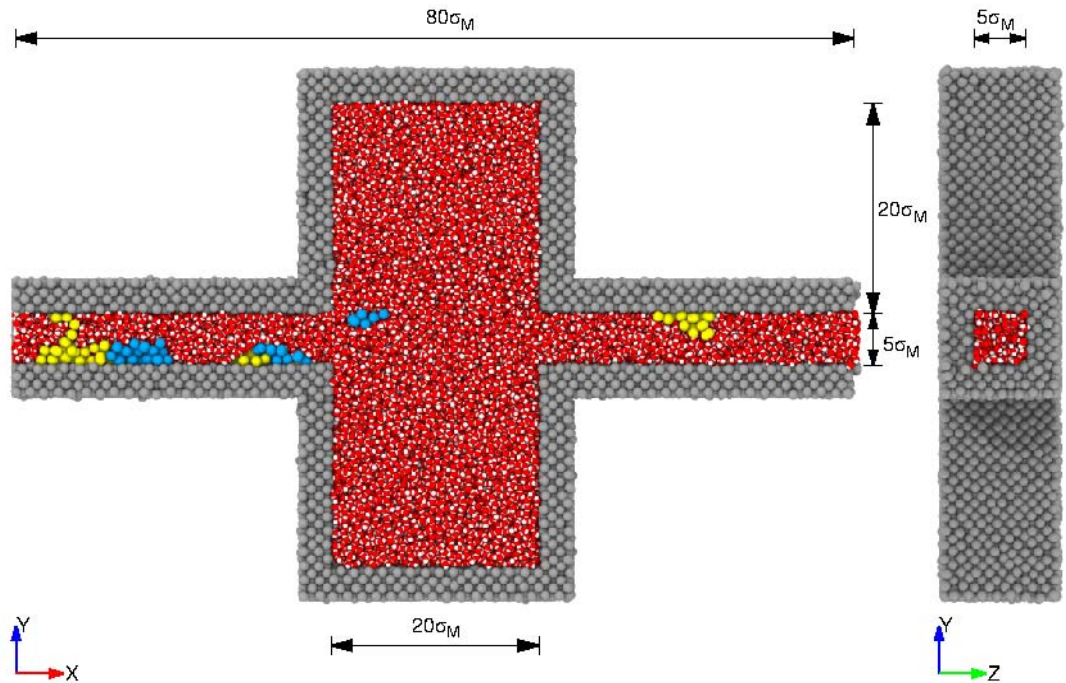

**FIGURE 1:** Dimensions of nanochannel in terms of $\sigma_M$ : Front view with front wall deleted and the side view. In the front view, polymers in blue colour is linear and the ones in yellow are star polymers. Water molecules are shown in red (Oxygen) and white (Hydrogen).

   In this study, classical Molecular Dynamics (MD) simulations are performed on aqueous solutions of polymers, with polymeric chains having various wettabilities and structures. We consider flow of polymer solutions, with linear and star configurations, confined between narrow channels with interactions tuneable via variation of Lennard-Jones potential parameters[28]. This type of interaction causes the polymer to get adsorbed on the walls. This aspect, in spite of its practical implications, has not been considered in any of the earlier studies. The surface effect of the confinement will be more pronounced in polymer dynamics when the dimensions of the channel are larger than the radius of gyration of the polymeric chains. Thus, the present study will also enable us to identify the locations in a channel/matrix where polymers will aggregate because of wettability induced trapping in the vicinity of the confinement surface. Based on the wall wettability and polymer structure, we discuss the possible methods of segregation of polymers in solvated state. The preferential trapping, using tuneable wettability of the surface, can find major applications in ultrafiltration membranes and anti-fouling technology. Interestingly, we observe that polymer trapping occurs as a combination of interactions derived from wall wettability, chain flexibility and polymer structure. We propose a non-intuitive polymer sorting mechanism based on wettability-structure interaction, which could be used to separate star and linear polymers.

**RESULTS**

Hydrophobic polymers can be easily separated from hydrophilic polymers, as they form rigid globular structure in flow and never get trapped. Globular shape of hydrophobic polymers and less number of monomers in contact with wall contribute to their increased mobilities. For hydrophilic polymers, since the number of monomers which are in direct contact with the wall is more, their mobilities are less (see SI). However, it will be difficult to separate star and linear polymers by means of difference in wettability, as both undergo trapping in corrugated nanochannels. These polymers in a flow, irrespective of the structure, have almost identical mobilities, rendering separation in flow impossible. Therefore, we need to extract polymers of different structures, from a solvated flow, by using combined mobility and entropic interactions.

FIGURE 2 shows the variation of mean depth at which the polymer is trapped in the pit with wall wettability, as dictated by the relevant Lennard-Jones parameter $(\varepsilon_{WW})$, and the fraction of polymers trapped for linear polymers. The probability of trapping increases as the polymer-wall attraction increases. At low wettability, the wall will not have sufficient effect in pulling the polymers toward it in order to trap. Thus, the polymeric chains reside in the flow field of the channel where chain relaxation is possible for the simulated dimensions. In effect, the channel walls perceive the polymer chains and the solvent molecule as a uniform medium, and segregation based on solely wettability becomes impossible. With increasing wall wettability, the increased attraction between polymers and the wall aids the polymers to creep on the wall leading them to get trapped at the mouth of the pit. It is also seen that the mean depth, at which the polymers get trapped in the pit, depends upon the total number of polymers trapped inside that pit, apart from the polymer-wall attraction. With a polymer already trapped inside the pit, if a second polymer moves in, the former accommodates the latter by moving further inside the pit. In the case of linear polymers, as more chains come inside the pit, polymers which are already trapped will go further inside and will, in turn, increase the mean depth. At low wettability, because of their structural simplicity, linear polymers show only a weak tendency to get trapped.

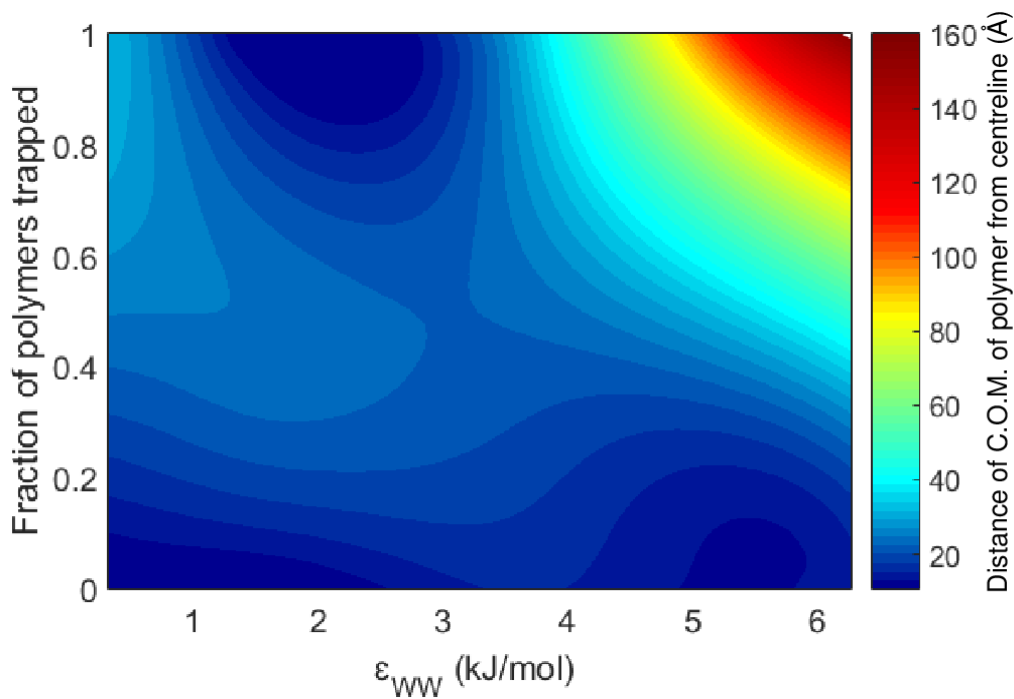

**FIGURE 2:** Dependence of mean depth of residence of center-of-mass (C.O.M) of linear polymers on fraction of polymers trapped inside a pit and wall wettability. Mean depth of residence increases as more number of polymers get trapped and as the wall wettability increases.

However, in the case of star polymers, the effect of polymer-wall wettability on trapping is more intricate than the case of linear polymers. This is induced by the structure and wettability interaction, which makes chain relaxation an important criterion even in large channels. When a star polymer gets trapped, it will try to unwind itself further and spread over the wall. This will be difficult and there is a possibility of entanglement, if it stays near to the fast moving streams in the channel. The polymeric chains try to avoid entanglement owing to the steric interactions in the solvated state. In order to achieve this, a trapped star polymer will move inside the pit along the wall. At low wettability, as more and more polymers get trapped in a pit, trapped polymers move further inside in order to avoid imminent entanglement with newly trapped ones and consequently, the mean depth increases (FIGURE 3). Thus, star polymers are observed to reside at lower pit levels as opposed to linear polymers residing at the mouth of the pit.

At higher wettabilities, because of the overly spread structure, it will be difficult for the trapped polymers to move further inside.[29] But, as more number of polymers moves in, the mean depth increases. Here, it is to be noted that the maximum depth at which star polymers reside is less than that of linear polymers, because of its complex structure. It is the effect of this uniform structure that compels the star polymers dive deeper into the pit at low wall wettabilities that also leads to the residence of star polymers near the mouth of the pit for higher wall wettabilities. While the behaviour at higher $\varepsilon_{ww}$ is similar to linear polymers, the increased trapping at lower wettabilities is solely contributed from the structure.

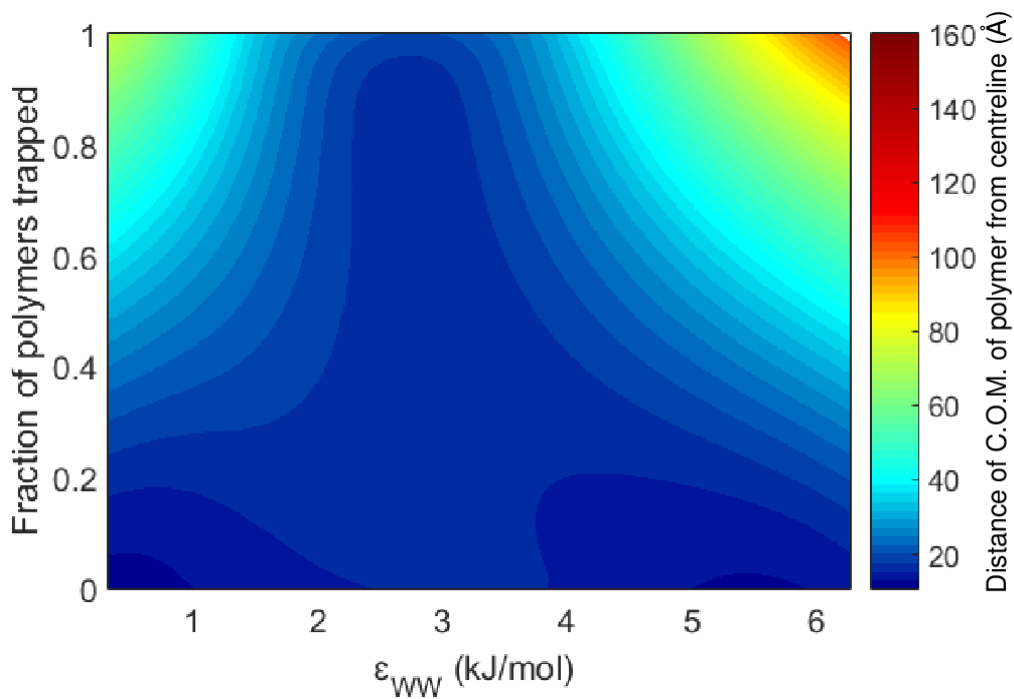

**FIGURE 3:** Dependence of mean depth of residence of C.O.M of star polymers on fraction of polymers trapped inside a pit and wall wettability. At low wettability, overly spread structure of star polymers on wall enables them to move into the pit as more number of polymers get trapped.

**DISCUSSIONS**

Star and linear polymers get trapped at both low and high wettabilities, albeit, the trapping mechanism being entirely contrary to each other. While the wettability based trapping at higher wettabilities is intuitive, the low wettability trapping must stem from the interaction of the polymer structure with the flow. Combined wettability and structural interactions alters the polymer mobility in the channels. The quantification of this mobility difference based on structure/conformation is indeed a good option to separate biopolymers during a flow through nanochannels. Both star and linear polymers are simulated in a channel of low wettability in order to demonstrate fractional separation of hydrophilic polymers based on structure, after getting trapped in the pit. As the results show, the difference in depth of residence for star and linear polymers reverses with the wettability. Hence, we use driving force as an independent variable in order to examine the cause of this reversal and devise ways to enhance this effect. It is seen that the star polymers are more mobile compared to linear polymers (FIGURE 4), because of the structure of star polymers and consequent difficulty in retarding them. In the case of star polymers, total force experienced by a polymer in the channel region is "centralised" in nature compared to linear polymers. It is interesting to note that this larger mobility for star polymers is more observable at intermediate driving forces. This non-monotonic dependence of polymer mobility difference on driving force can be considered in polymer separation techniques in order to separate different types of polymers having same degree of polymerization in aqueous medium, to achieve greater resolution at longer distances. This is because of the interplay between polymer-wall interaction and the opposing driving force.

In order to mimic the hydrophobic effect at lower wettabilities that causes non-intuitive trapping of star and linear polymers, we calculate the comparative mobilities at different driving forces. While the flexibility of the polymeric chains allows the polymers to relax into the pit, the opposing driving force makes them escape the attraction of pit wall and consequently, the trapped polymers, at larger driving force, equilibrate into the bulk solvent inside the pit. Moreover, at large driving force, polymers will not get sufficient time for equilibration as their mobility is high and independent of structure (FIGURE 4). In a previous study, researchers had noted a mobility independent of the degree of polymerisation, at large driving pressure differences for linear polymers.[30] Besides, in a pressure driven flow, the center-of-mass of the polymers tends to move away from the centerline of the channel, irrespective of whether the polymer is flexible or rigid. This occurs as a result of competition between hydrodynamic interactions of polymers with the wall and gradient in chain mobility across the channel. A tendency of the polymers to move away from wall and thus a shift in the peak of this distribution toward the centerline is also observed, as the driving force increases.[31,32] As a result, the adsorption of polymers on the wall surface decreases and consequently, wall wettability induced trapping will be low at large driving forces. Low wall wettabilities trigger similar interfacial gradient near the wall as induced by large driving force. Thus, the complex structure of star polymers under the influence of a velocity gradient would force them to move through the bulk and thus reside at greater depths for low wettability. Trapping at lower wettabilities occurs as a result of combined effect of interfacial depletion and chain relaxation. The water molecules and the polymeric chains tend to move away from the interface at lower wettabilities and under an applied driving force, considerable slippage would occur at the boundary. However, the presence of hydrophilic polymers would lead to chain elongation and as a result, entropic trapping due to chain relaxation of the polymer becomes more probable. The low polymer wall interaction ensures that the polymeric chains remain away from the wall and thus, the subsequent trajectory of the polymer inside the pit is guided by the steric interactions. Star polymers, due to their complex structure, try to avoid entanglement and thus entropic contributions in the flow tend to push it deeper as opposed to linear polymers.

Moreover, at low driving force, each polymer will get sufficient time to interact with wall, resulting in significant adsorption on it and consequently, there will be trapping in a transverse direction which renders mobilities of both the polymers low. But, as time proceeds, star polymers get trapped predominantly and move to greater depths, decreasing their mobilities. Thus, the wettability induced trapping at high wettability can be compared to the polymer dynamics at low driving forces. The inset of FIGURE 4 shows the track of center-of-mass of the both types of polymers in the nanochannels having contrasting wettabilities with low driving forces. Linear polymers are seen to dive deeper into the pit when the wettability of the channel walls is high (solid lines). However, when the channel walls have low wettabilities (dotted lines), because of chain relaxation, star polymers move deeper compared to linear polymers, thus achieving fractional separation of polymers.

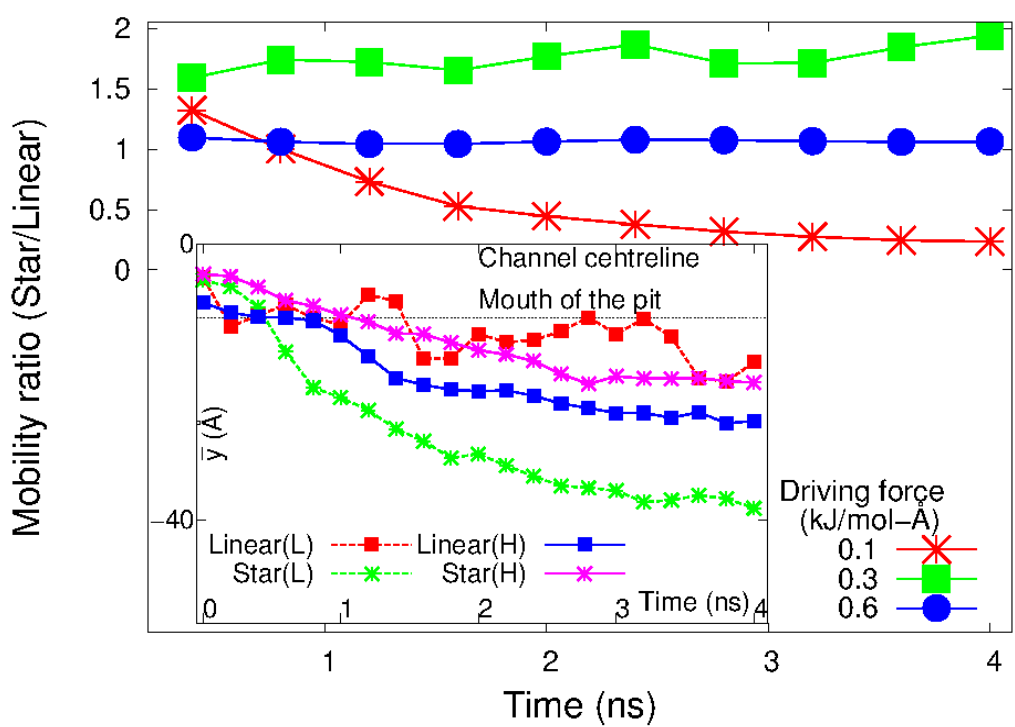

**FIGURE 4:** Mobility ratio for a flow of linear and star polymers. Mobilities of both the polymers are nearly same at high driving forces. At low driving forces, mobility of star polymers decline due to entanglement. At intermediate driving forces, due to the hydrophilic nature of the polymeric chains, star polymers tend to show higher mobilities as contributed from chain relaxation in the bulk solvent. Inset: Track of the center-of-mass $(\bar{y})$ of linear and star polymers in a nanochannel at low driving force. Dotted lines represent channels having low wettability (L) and solid lines indicate high wettability (H). In the former case the star polymers get preferentially trapped and move deeper into the pit. While, in the latter case, linear polymers prefer to move to a larger depth as compared to star polymers.

**CONCLUSIONS**

To summarize, we have studied the effect of polymer wettability on the confining wall of a corrugated nanochannel on the trapping of polymers, different in wettability and structure, in the presence of water, using MD simulations. Hydrophobic and hydrophilic polymers can be easily separated owing to their contrast in mobility that originates from their disparate conformations in solvated state. At higher wall wettabilities, hydrophilic polymers are increasingly trapped in the pit, as they creep along the wall in order to maintain their spread structure in the flow. While linear polymers are able to descend to greater depths in the pit owing to their structural simplicity, star polymers are segregated at the mouth of the pit. As opposed to this intuitive wettability based trapping at high wettabilities, trapping due to chain relaxation occurs at very low wall wettabilities. The depletion of the near wall molecules forces the polymers to relax into the pit, encouraging entropic trapping. However, the fractionation of star and linear polymers reverses at non-wetting walls. Star polymers, due to their complicated structure, are more prone to entropic trapping and recoiling in the bulk and thus attempt to dive deeper into the pit away from the flow field. As opposed to this, linear polymers are less effectively trapped in non-wetting channels and stay near to the mouth of the pit. This separation technique, being independent of the channel geometry, can be put to use in a wide variety of bio-filtration and industrial applications.

**ACKNOWLEDGEMENTS**

The authors gratefully acknowledge the financial support provided by the Indian Institute of Technology Kharagpur, India [Sanction Letter no.: IIT/SRIC/ATDC/CEM/2013-14/118, dated 19.12.2013].

**REFERENCES AND NOTES**

1. Wang, S.; Jing, B.; Zhu, Y. *J. Polym. Sci. Part B Polym. Phys.* **2014**, *52*, 85–103.
2. Fologea, D.; Gershow, M.; Ledden, B.; McNabb, D. S.; Golovchenko, J. a.; Li, J. *Nano Lett.* **2005**, *5*, 1905–1909.
3. Clarke, J.; Wu, H.-C.; Jayasinghe, L.; Patel, A.; Reid, S.; Bayley, H. *Nat. Nanotechnol.* **2009**, *4*, 265–270.
4. Min, S. K.; Kim, W. Y.; Cho, Y.; Kim, K. S. *Nat. Nanotechnol.* **2011**, *6*, 162–165.
5. Reisner, W.; Morton, K.; Riehn, R.; Wang, Y.; Yu, Z.; Rosen, M.; Sturm, J.; Chou, S.; Frey, E.; Austin, R. *Phys. Rev. Lett.* **2005**, *94*, 1–4.
6. Reisner, W.; Beech, J. P.; Larsen, N. B.; Flyvbjerg, H.; Kristensen, A.; Tegenfeldt, J. O. *Phys. Rev. Lett.* **2007**, *99*, 3–6.
7. Krishnan, M.; Mönch, I.; Schwille, P. *Nano Lett.* **2007**, *7*, 1270–1275.
8. Miller, J. B.; Hobbie, E. K. *J. Polym. Sci. Part B Polym. Phys.* **2013**, *51*, 1195–1208.
9. Barnes, M. D.; Mehta, A.; Kumar, P.; Sumpter, B. G.; Noid, D. W. *J. Polym. Sci. Part B Polym. Phys.* **2005**, *43*, 1571–1590.
10. Engelberg, I.; Kohn, J. *Biomaterials* **1991**, *12*, 292–304.

11. Ikada, Y. *Biomaterials* **1994**, *15*, 725–736.
12. Ulery, B. D.; Nair, L. S.; Laurencin, C. T. *J. Polym. Sci. Part B Polym. Phys.* **2011**, *49*, 832–864.
13. Overstreet, D. J.; Dutta, D.; Stabenfeldt, S. E.; Vernon, B. L. *J. Polym. Sci. Part B Polym. Phys.* **2012**, *50*, 881–903.
14. Han, J.; Turner, S.; Craighead, H. *Phys. Rev. Lett.* **1999**, *83*, 1688–1691.
15. Han, J.; Craighead, H. G. *Science (80-. ).* **2000**, *288*, 1026–1029.
16. Mahalik, J. P.; Yang, Y.; Deodhar, C.; Ankner, J. F.; Lokitz, B. S.; Kilbey, S. M.; Sumpter, B. G.; Kumar, R. *J. Polym. Sci. Part B Polym. Phys.* **2016**, n/a – n/a.
17. Rousseau, J.; Drouin, G.; Slater, G. W. *Phys. Rev. Lett.* **1997**, *79*, 1945–1948.
18. Slater, G. W.; Wu, S. Y. *Phys. Rev. Lett.* **1995**, *75*, 164.
19. Mikkelsen, M. B.; Reisner, W.; Flyvbjerg, H.; Kristensen, A. *Nano Lett.* **2011**, *11*, 1598–1602.
20. Han, J.; Craighead, H. G. **2002**, *74*, 394–401.
21. Turner, S. W. P.; Cabodi, M.; Craighead, H. G. *Phys. Rev. Lett.* **2002**, *88*, 128103.
22. Bakli, C.; Chakraborty, S. *J. Chem. Phys.* **2013**, *138*, 054504.
23. Bakli, C.; Chakraborty, S. *Appl. Phys. Lett.* **2012**, *101*, 153112.
24. Bakli, C.; Chakraborty, S. *Nano Lett.* **2015**, *15*, 7497–7502.
25. Erdtman, E.; Bohlén, M.; Ahlström, P.; Gkourmpis, T.; Berlin, M.; Andersson, T.; Bolton, K. *J. Polym. Sci. Part B Polym. Phys.* **2016**, *54*, 589–602.
26. Cao, Q.; Bachmann, M. *J. Polym. Sci. Part B Polym. Phys.* **2016**, n/a – n/a.
27. Akoum, R. Al; Vaulot, C.; Schwartz, D.; Hirn, M.-P.; Haidar, B. *J. Polym. Sci. Part B Polym. Phys.* **2010**, *48*, 2371–2378.
28. Allen, M. P.; Tildesley, D. J. *Computer Simulation of Fluids*; Clarendon Press, 1987.
29. Karagiovanaki, S.; Koutsioubas, A.; Spiliopoulos, N.; Anastassopoulos, D. L.; Vradis, A. A.; Toprakcioglu, C.; Siokou, A. E. *J. Polym. Sci. Part B Polym. Phys.* **2010**, *48*, 1676–1682.
30. Ollila, S. T. T.; Denniston, C.; Karttunen, M.; Ala-Nissila, T. *Phys. Rev. Lett.* **2014**, *112*, 118301.
31. Khare, R.; Graham, M. D.; De Pablo, J. J. *Phys. Rev. Lett.* **2006**, *96*, 9–12.
32. Park, J.; Bricker, J.; Butler, J. *Phys. Rev. E* **2007**, *76*, 1–4.